\documentclass[twocolumn]{aa}
\usepackage{graphicx}
\usepackage{txfonts}
\usepackage{natbib}
\bibpunct{(}{)}{;}{a}{}{,}
\begin{document}
   \title{An L0 dwarf companion in the brown dwarf desert, at 30~AU
\thanks{Based on observations made at Canada-France-Hawaii Telescope, 
operated by the National Council of Canada, the Centre National de la 
Recherche Scientifique de France and the University of Hawaii,
at the Observatoire de Haute Provence, operated by the Centre National de 
la Recherche Scientifique de France and at the  William Herschel Telescopes 
operated by the Isaac Newton Group at the Instituto de Astrof\'{\i}sica de 
Canarias}}
   \subtitle{}

   \author{T. Forveille\inst{1,2}   
      \and D. S\'egransan\inst{3}  
      \and P. Delorme\inst{1} 
      \and E.~L. Mart\'in\inst{4}
      \and X. Delfosse\inst{2}      
      \and J.~A. Acosta-Pulido\inst{4}
      \and J.-L. Beuzit\inst{2}
      \and A. Manchado\inst{4}      
      \and M. Mayor\inst{3}       
      \and C. Perrier\inst{2}
      \and S. Udry\inst{2} 
   }

   \offprints{T. Forveille : \email{Thierry.Forveille@cfht.hawaii.edu}}

   \institute{Canada-France-Hawaii Telescope Corporation, PO Box 1597, 
	       Kamuela, HI 96743, USA
         \and Laboratoire d'Astrophysique de Grenoble, BP 53X, F-38041 
	       Grenoble Cedex, France
         \and Observatoire de Gen\`eve, 51 Chemin des Maillettes, CH-1290, 
	       Switzerland 
         \and Instituto de Astrof\'{\i}sica de Canarias, C/ V\'ia L\'actea, 
	       s/n, E-38200 - La Laguna (Tenerife), Spain
             }

   \date{Received ; accepted }

   \abstract{
     We present the discovery of an L0 companion to
     the nearby M1.5 dwarf G~239-25, at a projected distance of 31~AU.
     It is the faintest companion discovered so far in our adaptive 
     optics survey of all known M dwarfs within 12~pc, and it lies 
     at the stellar/substellar limit. Given the assumed age of the primary
     star, the companion is likely an extremely low mass star.
     The long orbital period of G~239-25~AB ($\approx 100$ years) 
     precludes a direct mass determination, but the relatively wide angular
     separation will allow detailed analyses of its near infrared and visible
     spectra. 
   }

   \maketitle

\section{Introduction}
  Radial velocity surveys find that about 5\% of solar-type stars have
  planets within 4~AU with $M~\sin(i)$=0.25-13~$M_J$ 
  \cite[e.g.][]{2000prpl.conf.1285M}.
  The same surveys demonstrate, in contrast, that fewer than 1\% of these
  stars have more massive substellar companions ($M~\sin(i)$=13-80~$M_J$).
  This ``brown dwarf desert'' for close companions actually extends to 
  the lowest mass stellar companions, up to $\approx$100$M_J$, in keeping
  with the expectation that star formation does not care about the 
  substellar limit. 
  The frequency of more massive stellar companions in the same separation
  range is $\approx$10\% \citep{1991A&A...248..485D,2003A&A...397..159H}, 
  and this clear dichotomy of the mass
  distribution supports the idea that the stellar and planetary 
  companions to solar-type stars form through distinct channels.\\
  This ``desert'' stands in contrast to the relative abundance of 
  free-floating brown dwarfs in the field 
  \citep[e.g.][]{1999A&AS..135...41D,2003PASP..115..763C}, 
  and in young clusters down to very low masses 
  \citep[e.g.][]{1998A&A...336..490B, 2000Sci...290..103Z,2000ApJ...540.1016L}. 
  One open question is how far the brown dwarf desert extends beyond the 
  current 0-4~AU sensitivity range of the 
  radial velocity surveys. The prototype L and T dwarfs, GD~165B and
  Gl~229B, have  been found as companions to stars with initial masses
  within a factor of 2 of the Sun 
  \citep[][]{1992ApJ...386..260Z,1995Sci...270.1478O}, 
  and the limited statistics to date suggests that
  the brown dwarf desert may not exist beyond 250~AU from solar-type 
  stars \citep{2001ApJ...551L.163G} and that 1$\pm$1\% have companions
  in the 75-250~AU range \citep{2004AJ....127.2871M}.
  There is also good evidence
  that brown dwarfs are fairly common as close companions to other brown 
  dwarfs and to very low mass stars 
  \citep[e.g.][]{2003AJ....126.1526B, 2003ApJ...587..407C,2003ApJ...594..525M}.
  The situation at intermediate separations or intermediate mass ratios is 
  less clear, with probable brown dwarf found within 14~AU 
  and 47~AU of two solar analogs 
  \citep[][]{2002ApJ...571..519L,2002ApJ...567L.133P}  
  and  within 5 and 45~AU of two early M dwarfs 
  \citep[][]{1995Sci...270.1478O,2001astro.ph..6277N}.\\

  In this letter we present a new companion in
  this intermediate range around a nearby M1.5 dwarf, G~239-25.
  Section~2 discusses the observations and their analysis,
  while in Sect.~3 we examine the physical parameters of the binary 
  system. We then briefly discuss the population of L dwarfs companions 
  orbiting M dwarfs.

  \section{Observations \& data analysis}
\begin{figure}[h]
            \begin{center}  
                  \includegraphics[width=0.33\textwidth,angle=0]
                    {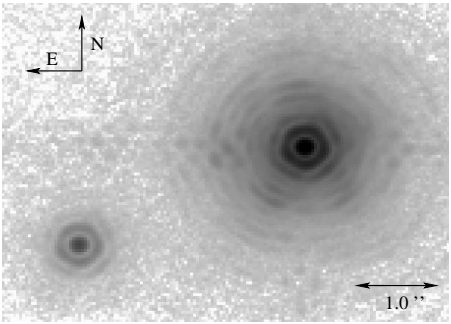}
              \end{center}%

   \caption{Adaptive optics image of the G~239-25~AB system in 
             Br$_{\gamma}$ in August 2001. 
    }
         \label{Fig-1}
   \end{figure}

   We observed G~239-25 in August 2001 with the 
   3.6-meter Canada-France-Hawaii-Telescope (CFHT), using the PUE`O adaptive 
   optics \citep{1998PASP..110..152R} and the KIR infrared camera, with
   observing and analysis procedures documented in \citet{astro-ph0106277}. 
   A possible companion to G~239-25 was easily seen in real time 
   (Fig.~\ref{Fig-1}), but we 
   needed follow-up observations to confirm its physical association.

   The probability of observing a background star within a few arcseconds
   of any given target is quite small at the $+$47$\deg$ galactic latitude of 
   G~239-25, but we have observed a sufficiently large parent sample that we
   could not {\it a priori} reject this possibility with extremely high 
   confidence.
   The two-band photometry from the J$_{cont}$ and Brackett~$\gamma$  
   ($Br_{\gamma}$)  filters 
   was also not strongly conclusive: the near-IR colours of very low mass 
   stars and brown dwarfs are only moderately distinctive, as opposed to
   their optical-to-IR color index; this general difficulty was compounded 
   here by observations through narrow-band filters for which we do not 
   have accurate colour transformations to standard photometric bands. \\

\begin{figure*}[t]

\vspace{-5truemm}\par\noindent
  \hbox to \textwidth{
     \parbox{0.5\textwidth}{
         \begin{center}
            \includegraphics[width=0.40\textwidth]
		{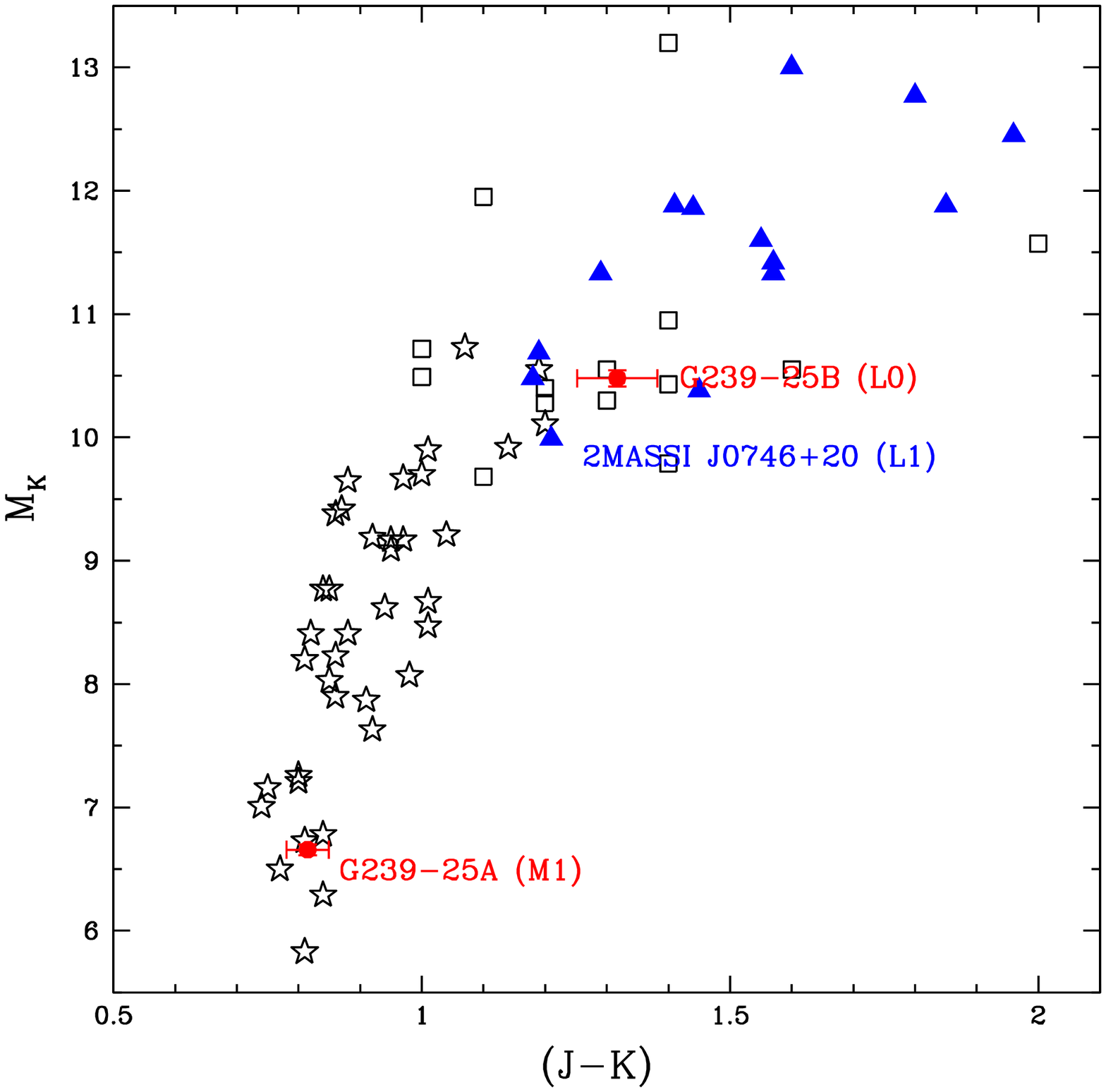}
         \end{center}
    }
     \hfil
     \parbox{0.5\textwidth}{
         \begin{center}
          \includegraphics[width=0.4\textwidth]
		{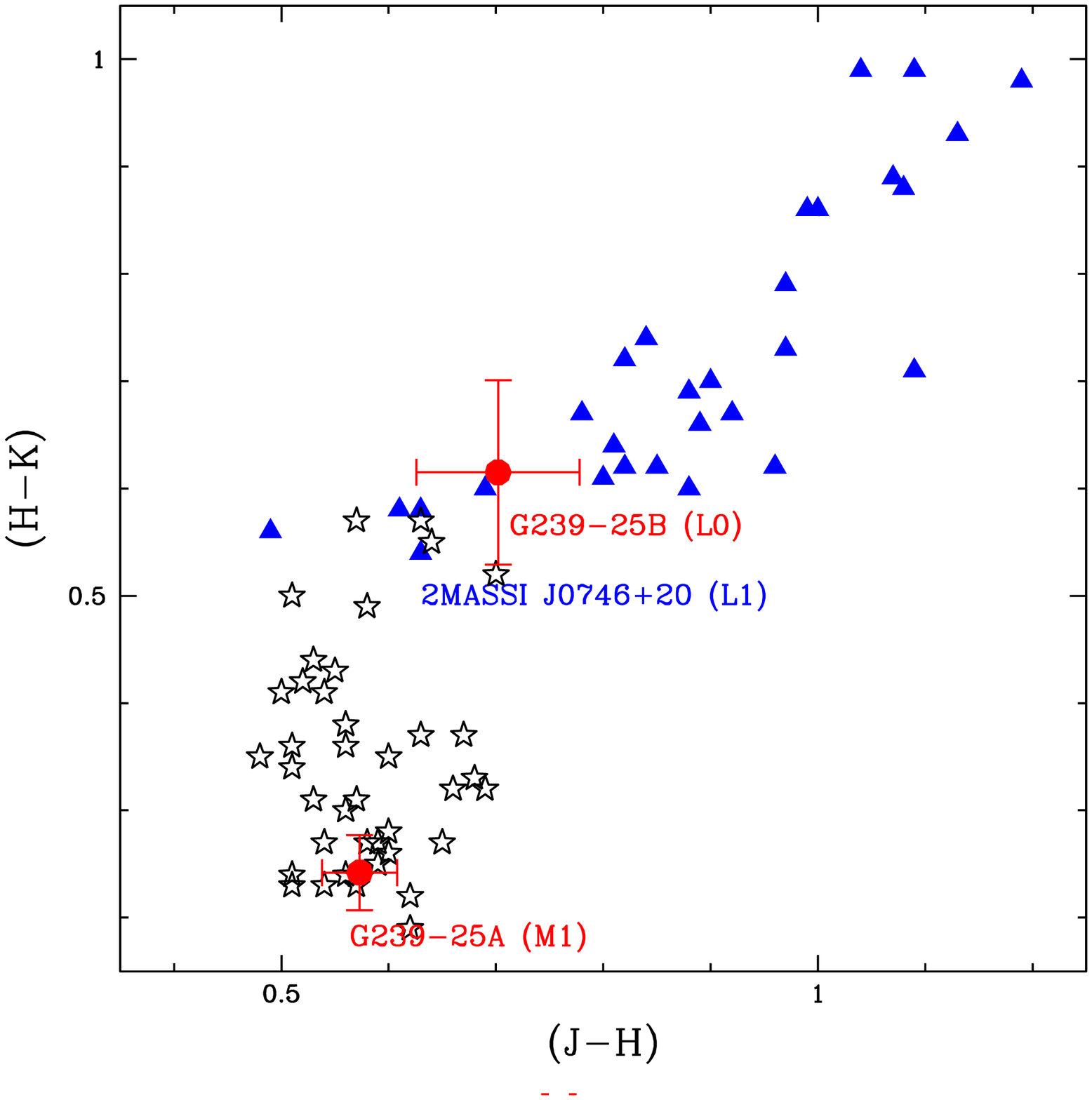}
         \end{center}
     }
 }  

    \caption{Absolute K magnitude vs. J-K (left) and color-color diagram of H-K vs. J-H (right) for M and L dwarfs.
    {\it Stars} correspond to M dwarfs  (\cite{2000ApJ...535..965L,2002ApJ...564..452L}) while  {\it filled triangles} 
    (\cite{2002ApJ...564..452L}) and  {\it open squares} (\cite{2004A&A...416L..17K}) correspond to L dwarfs. 
     G~239-25~A and B are represented by  {\it filled circles} with error bars.} 
        \label{Fig-2}
  \end{figure*}

   Like most nearby stars G~239-25 fortunately has a high proper motion 
   ($\mu_{\alpha}$=$-310.62\,\,mas/yr$, $\mu_{\delta}$=$-59.23\,\,mas/yr$),
   and it moves relative to background stars by $9~pixel/yr$.
   Common proper motion is thus easily tested after just a few months,
   and we have now reobserved G~239-25 on several occasions. We also found 
   from the {\it HST} archive that G~239-25 was observed in 
   November 1998 with NICMOS, which doubles the time span of the
   observations. For each PUE`O observing run, we determine the detector 
   scale and orientation by observing either the inner part of the Orion 
   Trapezium or some wide and distant HIPPARCOS binaries. We typically 
   achieve a pixel scale accuracy of $0.01\,mas/pixel$ and determine the 
   orientation of the detector with  $0.08\,deg$ precision. 
   Table {\ref{tab-1}} displays the parameters of the binary system at each 
   observed epoch. 
  \begin{table}
      \caption[]{Adaptive optics and HST measurements of G~239-25. Plate scale and orientation error from HST have not been taken into account.}
         \label{tab-1}

         \begin{tabular}{|cc|ccc|}
            \hline
            Date      &  Filter / $\lambda_{c}$& $\rho$   & P.A.     & $\Delta$mag\\
                      &          ($\mu m$)     & (arcsec) & (degree) & \\
            \hline     
              {\small 07Nov1998} & {\small F110W      /1.128} & {\small 3.028 $\pm$ 0.010} & {\small 114.13$\pm$0.18}&{\small   4.40$\pm$0.13} \\ 
              {\small 07Nov1998} & {\small F180M      /1.797} & {\small 3.024 $\pm$ 0.001} & {\small 114.15$\pm$0.03}&{\small   4.10$\pm$0.02} \\
              {\small 07Nov1998} & {\small F207M      /2.082} & {\small 3.025 $\pm$ 0.001} & {\small 114.17$\pm$0.02}&{\small   3.94$\pm$0.01} \\ 
              {\small 07Nov1998} & {\small F222M      /2.218} & {\small 3.023 $\pm$ 0.001} & {\small 114.22$\pm$0.02}&{\small   3.76$\pm$0.01} \\
               \hline
              {\small 06Aug2001} & {\small J$_{cont}$ /1.207}  & {\small 2.922 $\pm$ 0.006} & {\small 111.0 $\pm$0.2} &  {\small 4.21$\pm$0.04}\\
              {\small 06Aug2001} & {\small Br$\gamma$ /2.166}  & {\small 2.906 $\pm$ 0.006} & {\small 111.0 $\pm$0.2} &  {\small 3.83$\pm$0.04}\\
              {\small 28Apr2002} & {\small Br$\gamma$ /2.166}  & {\small 2.89  $\pm$ 0.01 } & {\small 110.1 $\pm$0.2} &  {\small 3.85$\pm$0.07}\\ 
              {\small 25Jun2002} & {\small H2         /2.122}  & {\small 2.87  $\pm$ 0.01 } & {\small 110.2 $\pm$0.2} &  {\small 3.91$\pm$0.05}\\
              {\small 05Apr2004} & {\small H2         /2.122}  & {\small 2.810 $\pm$ 0.008} & {\small 108.4 $\pm$0.3} &  {\small 3.83$\pm$0.04}\\ 
            \hline     
         \end{tabular}
   \end{table}

   \begin{figure}[h]
            \begin{center}  
                  \includegraphics[width=0.40\textwidth,angle=0]
                    {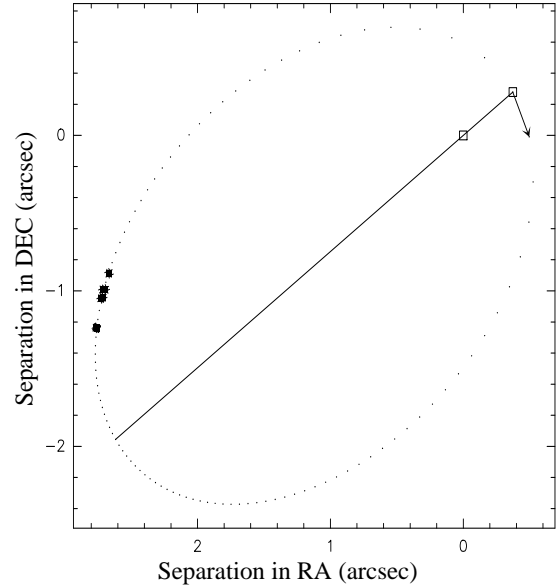}
              \end{center}%

   \caption{Tentative visual orbit of the G239-25 pair (North is up and East is left). The orbit is highly indeterminate, and should not be relied upon for more than observation planning over hte next few years. The predicted separations and positions angle on January 1$^{st}$ are (2.76,~107.0) for 2005,  (2.71,~105.7) for 2006,   (2.66,~104.4) for 2007 and (2.59,~102.9) for 2008. 
    }
         \label{Fig-3}
   \end{figure}

\begin{figure}[h]
            \begin{center}  
                  \includegraphics[width=0.40\textwidth]
                    {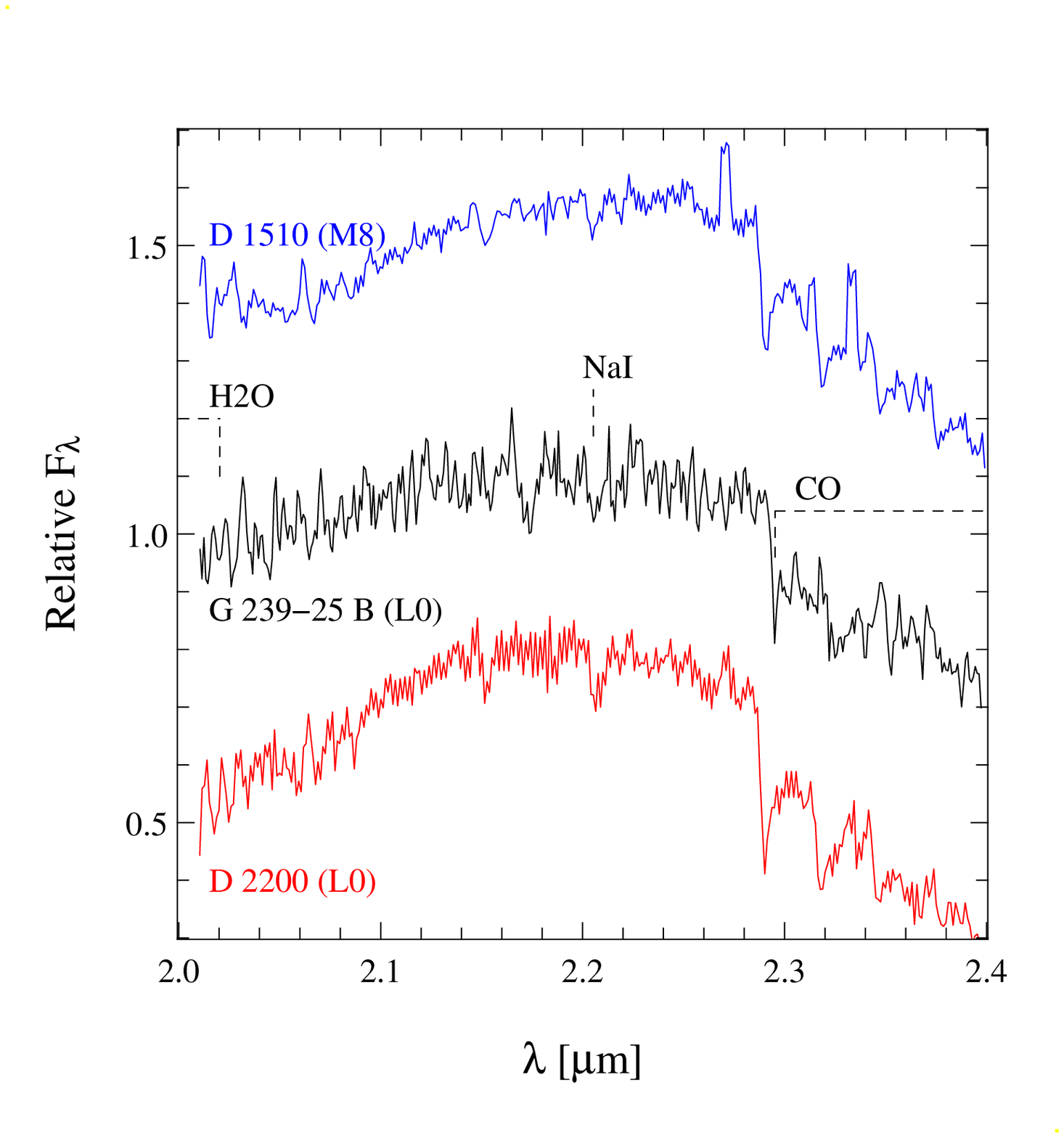}
              \end{center}%

   \caption{Near infrared spectrum of G~239-25~B (March 2004) compared 
             to Denis 1510 (M8)  and  to Denis 2200 (L0), \cite{2004A&A...416L..17K}.
             The estimation of the spectral type of G239-25~B is based on the depth of the CO absorbtion band 
             which is very similar to Denis 2200.
    }
         \label{Fig-4}
   \end{figure}

   The relative displacement of the two components over 6~years is
   ${\Delta}_{\alpha}$=-0.097'' and ${\Delta}{\delta}$=+0.351'',
   5~times smaller than the proper motion and in a different
   direction. It is, on the other hand, consistent with the 
   expected keplerian motion of the system and shows incipient 
   curvature (see Fig.~\ref{Fig-3} for a tentative orbit). \\
   Due the intrinsic dispersion of spectral type vs colour or absolute 
   magnitude \citep[e.g.][]{2004AJ....127.3553K}, compounded
   here by non-standard filters, our photometry constrains the spectral 
   class of the companion rather loosely, to an M9-L3 interval 
   (see Fig~\ref{Fig-2}).
   We therefore obtained infrared spectra of both components of G~239-25
   during the commissioning of LIRIS, the new near-infrared spectrograph 
   of the William Herschel Telescope 
   \citep{2000SPIE.4008.1162M,2003INGN....7...15A}. The 
   spectra have a resolution of R~$\approx$~700 and cover the 1 to 2.4${\mu}m$
   wavelength range. A spectral index analysis of the G~239-25~B spectrum 
   gives a spectral type of L0$\pm$1  based on the depth of the CO absorbtion band. 
   Fig.~\ref{Fig-4} demonstrates that it is visually very similar to our 
   L0 template. We will present a detailed quantitative analysis of this 
   spectrum in a forthcoming paper. 

  \section{Physical parameters}
  \begin{table*}
   \center
   \tabcolsep 0.1cm
      \caption[]{Photometry and color of G~239-25 using : 
                 $^{a}$ \citep{2002AJ....123.2806R}, 
                 $^{b}$\citep{2003yCat.2246....0C},
                 $^{c}$ this paper, CFHT data ,
                 $^{d}$ this paper, HST data
                 $^{e}$ the parallax value for G~239-25 is from a note to the on-line HIPPARCOS catalog and supersedes the printed parallax. }
         \begin{tabular}{|ccccccccccc|}
            \hline
            Name        &$\pi$       &$\mu_{\alpha}$&$\mu_{\beta}$&Spectral &        V         & I         & J         & H         & K         & J-K\\
                        & (mas)      &(mas/yr)      &(mas/yr)     &type     &              &          &          &          &          & \\
            \hline

             G~239-25~A & 92.62$\pm$1.52$^{e}$ &  -302.09$\pm$1.58 &  -33.35$\pm$1.86&M1.5&10.83$^{a}$&8.59$^{a}$&7.306$\pm$0.024$^{b}$&6.733$\pm$0.026$^{b}$&6.491$\pm$0.024$^{b}$&0.815$^{b}$\\             
             G~239-25~B & 92.62$\pm$1.52$^{e}$ &  -        & -                        &L0$\pm$1&  -        & -   &11.51$\pm$0.03$^{c}$          &10.83$\pm$0.03$^{d}$          &10.33$\pm$0.07$^{c,d}$&1.175$^{c,d}$ \\
            \hline
         \end{tabular}
         \label{tab-2}
   \end{table*}

  G~239-25~A is an M1.5 dwarf with $V$=$10.83$ and $K$=$6.49$ 
  \citep{2002AJ....123.3409H}. It belongs to the immediate solar 
  neighborhood, at a distance of only 10.8~pc ($\pi=92.62\pm1.52$~{\it mas},
  Table \ref{tab-2}).
  Visible spectra taken at Observatoire de Haute Provence with 
  the ELODIE Echelle spectrograph \citep{1996A&AS..119..373B} show 
  no signs of chromospheric activity ($H_{\alpha}$, $H_{\beta}$, or CaII
  H and K emission), and they put a low upper limits of 
  $vsin(i)<3~km/s$  to the projected rotational velocity. It 
  is thus not very young, but its kinematics ((U,V,W)=(22, 32, 14)~km/s)
  suggest that G~239-25 belongs to the young disk and is not an old star
  either.
  The ROSAT detection of X-ray emission and flares on G~239-25 
  \citep{1999A&AS..135..319H, 2003A&A...403..247F} further
  supports a moderately young age for that system. We tentatively adopt
  an age of 1 to 3~Gyr for G~239-25.

  The 2MASS infrared photometry \citep{2003yCat.2246....0C} and the 
  parallax result in absolute magnitudes of $M_{J}$=$7.14$, $M_{H}$=$6.57$, 
  $M_{K}$=$6.32$ for G~239-25~A.
  Theoretical and empirical mass-luminosity relations 
  \citep{1998A&A...337..403B,2000A&A...364..217D}
  both lead to a $0.4M_{\odot}$ mass for the primary, for any age between
  200~Myr and 15~Gyrs \citep{1998A&A...337..403B}.
  The mass derived for the companion is, unsurprisingly, very sensitive
  to the age. 
  The absolute K~band magnitude from the evolutionary models make G239-25~B
  a brown dwarf if it is younger than 500~Myr
  \citep{2000ApJ...542..464C}, so for the adopted age range it would be 
  stellar. The uncertainties in the models, the photometry, and most
  seriously the age, are however easily sufficient to make it a brown dwarf.

\section{Discussion}
The new L~dwarf was discovered while observing a volume-limited 
sample, and is amenable to a crude statistical analysis
of the frequency of brown dwarf companions around early M~dwarfs. The parent
sample of the observing program \citep{2001astro.ph..6277N} comprises the 
$\approx$450~M dwarfs within 12~pc, of which $\approx$300 are M4.5 and 
earlier. To date we have observed about 250 of those, and we would have 
easily detected G~239-25~B in every case. 
Our only other detection of a 
faint companion beyond 10~AU of its primary is Gl~229B 
\citep{1995Sci...270.1478O}, 
the prototype of the T dwarf spectral class. 
Its projected distance from its M1.5 primary is 44~AU, quite similar 
to G~239-25~B. The two detections in a sample of 250 suggest that 
$\approx$1\% of early M dwarfs have L and T dwarf companions orbiting 
between 10 and 50~AU. 

\vspace{0.25cm}
{\it Note added in proofs:} An independant discovery of G~239-25~B has been announced by
Golimowski et al. (AJ in press, astro-ph0406664) during the refereeing of the present
letter. Their conclusions are consistent with the photometric part of our analysis.

\begin{acknowledgements}
We thank the anonymous referee for a detailed and constructive report,
which improved the presentation of this work.
This research has made use of the Simbad database operated at CDS 
(Strasbourg, France), as well as of the HST Archived Exposures Catalog (2004).
This publication also makes use of data products from the 
Two Micron All Sky Survey, which is a joint project of the University of 
Massachusetts and the Infrared Processing and Analysis Center/California 
Institute of Technology, funded by the National Aeronautics and Space 
Administration and the National Science Foundation.
D. S\'egransan  acknowledges the support of the \emph{Fonds National de la 
Recherche Scientifique Suisse}.
\end{acknowledgements}

\bibliographystyle{aa}

\bibliography{dsegransan.bib}

\begin{thebibliography}{35}
\expandafter\ifx\csname natexlab\endcsname\relax\def\natexlab#1{#1}\fi

\bibitem[{{Acosta Pulido} {et~al.}(2003){Acosta Pulido}, {Ballesteros},
  {Barreto}, {Barreto}, {Cadavid}, {Carrillo}, {Correa}, {Delgado},
  {Dom{\'{\i}}nguez-Tagle}, {Hern{\' a}ndez}, {L{\' o}pez}, {Manescau},
  {Manchado}, {Moreno}, {Olives}, {Peraza}, {Prada}, {Redondo}, {S{\' a}nchez},
  {Sosa}, \& {Tenegi}}]{2003INGN....7...15A}
{Acosta Pulido}, J.~A., {Ballesteros}, E., {Barreto}, M., {et~al.} 2003, The
  Newsletter of the Isaac Newton Group of Telescopes (ING Newsl.), issue no.~7,
  p.~15-16., 7, 15

\bibitem[{{Baraffe} {et~al.}(1998){Baraffe}, {Chabrier}, {Allard}, \&
  {Hauschildt}}]{1998A&A...337..403B}
{Baraffe}, I., {Chabrier}, G., {Allard}, F., \& {Hauschildt}, P.~H. 1998, \aap,
  337, 403

\bibitem[{{Baranne} {et~al.}(1996){Baranne}, {Queloz}, {Mayor}, {Adrianzyk},
  {Knispel}, {Kohler}, {Lacroix}, {Meunier}, {Rimbaud}, \&
  {Vin}}]{1996A&AS..119..373B}
{Baranne}, A., {Queloz}, D., {Mayor}, M., {et~al.} 1996, \aaps, 119, 373

\bibitem[{{Beuzit} {et~al.}(2004{\natexlab{a}}){Beuzit}, {}, \&
  {Mayor}}]{astro-ph0106277}
{Beuzit}, J.-L., {}, W.~D., \& {Mayor}, M. 2004{\natexlab{a}}, \aap

\bibitem[{{Beuzit} {et~al.}(2004{\natexlab{b}}){Beuzit}, {S\'{e}gransan},
  {Forveille}, {Udry}, {Delfosse}, {Mayor}, {Perrier}, {Hainaut}, {Roddier},
  {Roddier}, \& {Mart{\'{\i}}n}}]{2001astro.ph..6277N}
{Beuzit}, J.~L., {S\'{e}gransan}, D., {Forveille}, T., {et~al.}
  2004{\natexlab{b}}, accepted by A\&A, astro-ph/0106277

\bibitem[{{Bouvier} {et~al.}(1998){Bouvier}, {Stauffer}, {Martin}, {Barrado y
  Navascues}, {Wallace}, \& {Bejar}}]{1998A&A...336..490B}
{Bouvier}, J., {Stauffer}, J.~R., {Martin}, E.~L., {et~al.} 1998, \aap, 336,
  490

\bibitem[{{Bouy} {et~al.}(2003){Bouy}, {Brandner}, {Mart{\'{\i}}n}, {Delfosse},
  {Allard}, \& {Basri}}]{2003AJ....126.1526B}
{Bouy}, H., {Brandner}, W., {Mart{\'{\i}}n}, E.~L., {et~al.} 2003, \aj, 126,
  1526

\bibitem[{{Chabrier}(2003)}]{2003PASP..115..763C}
{Chabrier}, G. 2003, \pasp, 115, 763

\bibitem[{{Chabrier} {et~al.}(2000){Chabrier}, {Baraffe}, {Allard}, \&
  {Hauschildt}}]{2000ApJ...542..464C}
{Chabrier}, G., {Baraffe}, I., {Allard}, F., \& {Hauschildt}, P. 2000, \apj,
  542, 464

\bibitem[{{Close} {et~al.}(2003){Close}, {Siegler}, {Freed}, \&
  {Biller}}]{2003ApJ...587..407C}
{Close}, L.~M., {Siegler}, N., {Freed}, M., \& {Biller}, B. 2003, \apj, 587,
  407

\bibitem[{{Cutri} {et~al.}(2003){Cutri}, {Skrutskie}, {van Dyk}, {Beichman},
  {Carpenter}, {Chester}, {Cambresy}, {Evans}, {Fowler}, {Gizis}, {Howard},
  {Huchra}, {Jarrett}, {Kopan}, {Kirkpatrick}, {Light}, {Marsh}, {McCallon},
  {Schneider}, {Stiening}, {Sykes}, {Weinberg}, {Wheaton}, {Wheelock}, \&
  {Zacarias}}]{2003yCat.2246....0C}
{Cutri}, R.~M., {Skrutskie}, M.~F., {van Dyk}, S., {et~al.} 2003, VizieR Online
  Data Catalog, 2246, 0

\bibitem[{{Delfosse} {et~al.}(2000){Delfosse}, {Forveille}, {S{\' e}gransan},
  {Beuzit}, {Udry}, {Perrier}, \& {Mayor}}]{2000A&A...364..217D}
{Delfosse}, X., {Forveille}, T., {S{\' e}gransan}, D., {et~al.} 2000, \aap,
  364, 217

\bibitem[{{Delfosse} {et~al.}(1999){Delfosse}, {Tinney}, {Forveille},
  {Epchtein}, {Borsenberger}, {Fouqu{\' e}}, {Kimeswenger}, \& {Tiph{\`
  e}ne}}]{1999A&AS..135...41D}
{Delfosse}, X., {Tinney}, C.~G., {Forveille}, T., {et~al.} 1999, \aaps, 135, 41

\bibitem[{{Duquennoy} \& {Mayor}(1991)}]{1991A&A...248..485D}
{Duquennoy}, A. \& {Mayor}, M. 1991, \aap, 248, 485

\bibitem[{{Fuhrmeister} \& {Schmitt}(2003)}]{2003A&A...403..247F}
{Fuhrmeister}, B. \& {Schmitt}, J.~H.~M.~M. 2003, \aap, 403, 247

\bibitem[{{Gizis} {et~al.}(2001){Gizis}, {Kirkpatrick}, {Burgasser}, {Reid},
  {Monet}, {Liebert}, \& {Wilson}}]{2001ApJ...551L.163G}
{Gizis}, J.~E., {Kirkpatrick}, J.~D., {Burgasser}, A., {et~al.} 2001, \apjl,
  551, L163

\bibitem[{{H{\" u}nsch} {et~al.}(1999){H{\" u}nsch}, {Schmitt}, {Sterzik}, \&
  {Voges}}]{1999A&AS..135..319H}
{H{\" u}nsch}, M., {Schmitt}, J.~H.~M.~M., {Sterzik}, M.~F., \& {Voges}, W.
  1999, \aaps, 135, 319

\bibitem[{{Halbwachs} {et~al.}(2003){Halbwachs}, {Mayor}, {Udry}, \&
  {Arenou}}]{2003A&A...397..159H}
{Halbwachs}, J.~L., {Mayor}, M., {Udry}, S., \& {Arenou}, F. 2003, \aap, 397,
  159

\bibitem[{{Hawley} {et~al.}(2002){Hawley}, {Covey}, {Knapp}, {Golimowski},
  {Fan}, {Anderson}, {Gunn}, {Harris}, {Ivezi{\' c}}, {Long}, {Lupton},
  {McGehee}, {Narayanan}, {Peng}, {Schlegel}, {Schneider}, {Spahn}, {Strauss},
  {Szkody}, {Tsvetanov}, {Walkowicz}, {Brinkmann}, {Harvanek}, {Hennessy},
  {Kleinman}, {Krzesinski}, {Long}, {Neilsen}, {Newman}, {Nitta}, {Snedden}, \&
  {York}}]{2002AJ....123.3409H}
{Hawley}, S.~L., {Covey}, K.~R., {Knapp}, G.~R., {et~al.} 2002, \aj, 123, 3409

\bibitem[{{Kendall} {et~al.}(2004){Kendall}, {Delfosse}, {Mart{\'{\i}}n}, \&
  {Forveille}}]{2004A&A...416L..17K}
{Kendall}, T.~R., {Delfosse}, X., {Mart{\'{\i}}n}, E.~L., \& {Forveille}, T.
  2004, \aap, 416, L17

\bibitem[{{Knapp} {et~al.}(2004){Knapp}, {Leggett}, {Fan}, {Marley}, {Geballe},
  {Golimowski}, {Finkbeiner}, {Gunn}, {Hennawi}, {Ivezi{\' c}}, {Lupton},
  {Schlegel}, {Strauss}, {Tsvetanov}, {Chiu}, {Hoversten}, {Glazebrook},
  {Zheng}, {Hendrickson}, {Williams}, {Uomoto}, {Vrba}, {Henden}, {Luginbuhl},
  {Guetter}, {Munn}, {Canzian}, {Schneider}, \&
  {Brinkmann}}]{2004AJ....127.3553K}
{Knapp}, G.~R., {Leggett}, S.~K., {Fan}, X., {et~al.} 2004, \aj, 127, 3553

\bibitem[{{Leggett} {et~al.}(2000){Leggett}, {Allard}, {Dahn}, {Hauschildt},
  {Kerr}, \& {Rayner}}]{2000ApJ...535..965L}
{Leggett}, S.~K., {Allard}, F., {Dahn}, C., {et~al.} 2000, \apj, 535, 965

\bibitem[{{Leggett} {et~al.}(2002){Leggett}, {Golimowski}, {Fan}, {Geballe},
  {Knapp}, {Brinkmann}, {Csabai}, {Gunn}, {Hawley}, {Henry}, {Hindsley},
  {Ivezi{\' c}}, {Lupton}, {Pier}, {Schneider}, {Smith}, {Strauss}, {Uomoto},
  \& {York}}]{2002ApJ...564..452L}
{Leggett}, S.~K., {Golimowski}, D.~A., {Fan}, X., {et~al.} 2002, \apj, 564, 452

\bibitem[{{Liu} {et~al.}(2002){Liu}, {Fischer}, {Graham}, {Lloyd}, {Marcy}, \&
  {Butler}}]{2002ApJ...571..519L}
{Liu}, M.~C., {Fischer}, D.~A., {Graham}, J.~R., {et~al.} 2002, \apj, 571, 519

\bibitem[{{Luhman} {et~al.}(2000){Luhman}, {Rieke}, {Young}, {Cotera}, {Chen},
  {Rieke}, {Schneider}, \& {Thompson}}]{2000ApJ...540.1016L}
{Luhman}, K.~L., {Rieke}, G.~H., {Young}, E.~T., {et~al.} 2000, \apj, 540, 1016

\bibitem[{{Manchado} {et~al.}(2000){Manchado}, {Barreto}, {Acosta-Pulido},
  {Prada}, {Dominguez-Tagle}, {Correa}, {Fragoso-Lopez}, {Fuentes}, {Iserte},
  {Joven-Alvarez}, {Lopez}, {Manescau}, {Moreno-Arce}, {Padron}, {Rasilla},
  {Redondo}, {Sanchez}, {Sosa}, {Atad-Ettedgui}, \&
  {Hastings}}]{2000SPIE.4008.1162M}
{Manchado}, A., {Barreto}, M., {Acosta-Pulido}, J., {et~al.} 2000, in Proc.
  SPIE Vol. 4008, p. 1162-1171, Optical and IR Telescope Instrumentation and
  Detectors, Masanori Iye; Alan F. Moorwood; Eds., 1162--1171

\bibitem[{{Marcy} {et~al.}(2000){Marcy}, {Cochran}, \&
  {Mayor}}]{2000prpl.conf.1285M}
{Marcy}, G.~W., {Cochran}, W.~D., \& {Mayor}, M. 2000, Protostars and Planets
  IV, 1285

\bibitem[{{Mart{\'{\i}}n} {et~al.}(2003){Mart{\'{\i}}n}, {Barrado y Navascu{\'
  e}s}, {Baraffe}, {Bouy}, \& {Dahm}}]{2003ApJ...594..525M}
{Mart{\'{\i}}n}, E.~L., {Barrado y Navascu{\' e}s}, D., {Baraffe}, I., {Bouy},
  H., \& {Dahm}, S. 2003, \apj, 594, 525

\bibitem[{{McCarthy} \& {Zuckerman}(2004)}]{2004AJ....127.2871M}
{McCarthy}, C. \& {Zuckerman}, B. 2004, \aj, 127, 2871

\bibitem[{{Oppenheimer} {et~al.}(1995){Oppenheimer}, {Kulkarni}, {Matthews}, \&
  {Nakajima}}]{1995Sci...270.1478O}
{Oppenheimer}, B.~R., {Kulkarni}, S.~R., {Matthews}, K., \& {Nakajima}, T.
  1995, Science, 270, 1478

\bibitem[{{Potter} {et~al.}(2002){Potter}, {Mart{\'{\i}}n}, {Cushing},
  {Baudoz}, {Brandner}, {Guyon}, \& {Neuh{\" a}user}}]{2002ApJ...567L.133P}
{Potter}, D., {Mart{\'{\i}}n}, E.~L., {Cushing}, M.~C., {et~al.} 2002, \apjl,
  567, L133

\bibitem[{{Reid} \& {Cruz}(2002)}]{2002AJ....123.2806R}
{Reid}, I.~N. \& {Cruz}, K.~L. 2002, \aj, 123, 2806

\bibitem[{{Rigaut} {et~al.}(1998){Rigaut}, {Salmon}, {Arsenault}, {Thomas},
  {Lai}, {Rouan}, {V{\' e}ran}, {Gigan}, {Crampton}, {Fletcher}, {Stilburn},
  {Boyer}, \& {Jagourel}}]{1998PASP..110..152R}
{Rigaut}, F., {Salmon}, D., {Arsenault}, R., {et~al.} 1998, \pasp, 110, 152

\bibitem[{{Zapatero Osorio} {et~al.}(2000){Zapatero Osorio}, {B{\' e}jar},
  {Mart{\'{\i}}n}, {Rebolo}, {Barrado y Navascu{\' e}s}, {Bailer-Jones}, \&
  {Mundt}}]{2000Sci...290..103Z}
{Zapatero Osorio}, M.~R., {B{\' e}jar}, V.~J.~S., {Mart{\'{\i}}n}, E.~L.,
  {et~al.} 2000, Science, 290, 103

\bibitem[{{Zuckerman} \& {Becklin}(1992)}]{1992ApJ...386..260Z}
{Zuckerman}, B. \& {Becklin}, E.~E. 1992, \apj, 386, 260

\end{thebibliography}

\end{document}